\documentclass{aa}
\def\ll{\mbox{${\lambda\lambda}$}}
\def\lam{\mbox{${\lambda}$}}

\usepackage{graphicx}

\usepackage{txfonts}

\begin{document}
\title{Medium-resolution spectroscopy of galaxies with redshifts $2.3 < z < 3.5$
\thanks{Based on observations obtained with FORS2 at the ESO Very Large 
Telescope, Paranal, Chile (proposals 69.A-0105 and 71.A-0307). Send offprints 
requests to C. Tapken: tapken@mpia.de}}
\titlerunning{Medium-resolution spectroscopy of high-redshift galaxies}
\authorrunning{Mehlert et al.}

\author{D. Mehlert,\inst{1,2}  C. Tapken,\inst{1,2}
  I. Appenzeller,\inst{2} S. Noll,\inst{3} D. de
  Mello,\inst{4,5} \and T.M. Heckman\inst{6}}

\institute{Max-Planck-Institut f\"ur Astronomie, K{\"o}nigstuhl 17, 69117  Heidelberg, Germany\\             
\and
Landessternwarte Heidelberg, K{\"o}nigstuhl 12, 69117  Heidelberg, Germany
\and MPE Garching, Gie{\ss}enbachstrasse, 85748 Garching, Germany
\and Laboratory for Astronomy and Solar Physics, GSFC, Greenbelt,  MD 20771, USA
\and Department of Physics, Catholic University of America,  DC 20064, USA 
\and Department of Physics and Astronomy, Johns Hopkins University, Baltimore, MD 21218, USA}             
\date{}

\abstract{Using FORS2 at the ESO VLT we obtained medium resolution ($R \approx 2000$) spectra of 12 galaxies with  $2.37 \leq z \leq 3.40$ in the FORS Deep Field. Two individual spectra with good S/N and a composite of all 12 spectra were used to derive properties of the stellar and interstellar absorption lines of galaxies in this redshift range. Systematic differences between the individual spectra were found for the strength and profiles of the intrinsic interstellar lines. For eight
spectra with sufficient S/N we measured the `1370' and `1425' metallicity
indices.  From these indices we find  for our sample that galaxies at
  $z>3$ have lower mean metallicity than galaxies at 2.5$<z<$3. However there remain uncertainties concerning the absolute calibration of the metallicity tracers in use for high-redshift galaxies. Additional modeling will be needed to resolve these uncertainties.   
}   

\maketitle
\section{Introduction}
The formation and early evolution of galaxies is one of the key problems of present-day astronomy. Among the open questions are the evolution of the stellar populations and the chemical enrichment of galaxies at early cosmic epochs. Most known high-redshift galaxies, like the Lyman-break galaxies (LBGs) defined by Steidel et al. (\cite{STE96a}, \cite{STE96b}), show spectral energy distributions and spectral properties similar to those of  local starburst galaxies. Therefore, the $R_{23}$ method, developed by Pagel et al. (\cite{PAG79}) for local starburst galaxies and based on the nebular emission lines of $[$O\,II$]$, $[$O\,III$]$, and H$\beta$ has been applied to derive the oxygen abundance of the ionized gas in LBGs. Using this method Teplitz et al. (\cite{TEP00}), Kobulnicky \& Koo (\cite{KOB00}), and Pettini et al. (\cite{PET01}) derived LBG metallicities at $z \sim 3$ ranging typically between $\approx 0.1\,\mathrm{Z}_{\odot}$ and $\approx 0.9\,\mathrm{Z}_{\odot}$. But, because of the degeneracy in the $R_{23}$ calibration, the results are sometimes uncertain by up to one order of magnitude. More recently Shapley et al. (\cite{SHA04}) extended this work using emission lines of $[$N\,II$]$ and H$\alpha$ to show that the metallicity of optical bright galaxies at $z\approx 2$ may reach nearly solar values. For high-redshift galaxies the (restframe) optical nebular emission lines are shifted into the (observer's frame) NIR wavelength range which suffers from a strong sky background. Therefore $R_{23}$ studies of high-redshift galaxies have so far been restricted to relatively few bright objects. 

As shown, e.g., by Storchi-Bergmann et al. (\cite{STO94}) and Heckman et al. (\cite{HEC98}) information on the stellar and chemical composition of starburst galaxies can also be obtained from the UV spectra of such objects. For distant galaxies with $z \approx 3$ the restframe UV spectral range is shifted into the observer's frame optical spectral region which can be readily observed with groundbased optical telescopes. A common tool to derive information from such spectra is a comparison with synthetic (restframe UV) spectra, as published, e.g., by Leitherer et al. (\cite{LEI99}, \cite{LEI01}; hereafter SB99) which are calculated with different assumptions on the age, metallicity, and initial mass function. This method has been successfully applied to gravitationally lensed and amplified galaxies (Pettini et al. \cite{PET00}, \cite{PET02}; Mehlert et al. \cite{MEH01}; Frye et al. \cite{FRY02}) and to low-resolution composite spectra (Lowenthal et al. \cite{LOW97}; Steidel et al. \cite{STE01}; Steidel 
et al. \cite{STE03}; Shapley et al. \cite{SHA03}).
\begin{table*}
\caption{The subsample of FDF galaxies and their medium resolution spectra used in this study.}\label{tab:obs}
\vspace{0.5cm} 
\centerline{
\begin{tabular}{|c|c|c|c|c|c|c|c|c|}
\hline 
ID & $z$ & $m_R$ & $m_{K_s}$ & Grism & $\lambda$ range & SNR & Ly$\alpha$ EW & $\beta$ \\
 & & $[$mag$]$ & $[$mag$]$ & & $[$\AA{}$]$ & &$[$\AA{}$]$ & \\
\hline
FDF-0960 & 3.159 & 24.27 & 21.03 & 1200R & 5920.5 -- 7420.5 & 4.0 & 0.6 & -2.27 \\
FDF-1337 & 3.403 & 23.91 & 21.17 & 1400V & 4385.5 -- 5550.0 & 3.5 & 0.5 & -2.43 \\
         &       &       &       & 1200R & 5436.0 -- 6908.0 & 4.7 & & \\
FDF-3173 & 3.270 & 24.29 & 21.37 & 1200R & 5821.0 -- 7316.0 & 4.3 & 10.9 & -1.39 \\
FDF-3810 & 2.372 & 22.99 & 19.75 & 1400V & 4315.0 -- 5604.5 & 6.5 & 11.5 & -0.33 \\
FDF-4691 & 3.304 & 24.59 & 21.91 & 1400V & 4685.0 -- 5996.5 & 3.8 & -103 & -2.46 \\
FDF-5215 & 3.148 & 23.40 & 20.38 & 1400V & 4403.5 -- 5694.5 & 3.0 & -38.4 & -1.71 \\          &       &       &       & 1200R & 6089.0 -- 7600.0 & 4.0 & & \\  
FDF-5550 & 3.383 & 23.45 & 20.65 & 1400V & 4615.5 -- 5922.5 & 3.8 & 17.0 & -1.81 \\          &       &       &       & 1200R & 5761.5 -- 7259.5 & 6.5 & & \\ 
FDF-5903 & 2.774 & 22.61 & 20.19 & 1400V & 4457.0 -- 5750.0 & 12.3 & 9.3 & -1.18 \\
FDF-6024 & 2.372 & 22.31 & 19.86 & 1400V & 4778.5 -- 6093.5 & 10.3 & -8.5 & -0.99 \\
FDF-6063 & 3.397 & 22.87 & 21.00 & 1400V & 4778.5 -- 6086.5 & 4.2 & 18.5 & -2.00 \\
         &       &       &       & 1200R & 5597.5 -- 7075.0 & 5.8 & & \\
FDF-6934 & 2.445 & 23.12 & 20.17 & 1400V & 4553.5 -- 5851.0 & 4.2 & 22.1 & -0.26 \\
FDF-7539 & 3.287 & 23.79 & 21.25 & 1400V & 4809.0 -- 6123.5 & 3.5 & -12.1 & -1.74 \\         &       &       &       & 1200R & 5791.0 -- 7279.0 & 4.9 & & \\
\hline
\end{tabular}}                       
\end{table*}              

 In order to obtain further information on the properties and
  metallicities of UV-luminous starburst galaxies at high redshift  we
  obtained new medium-resolution ($R \approx 2000$) spectra of
  a subsample of high-redshift galaxies from the FORS Deep Field (FDF)
  spectroscopic catalog (Noll et al. \cite{NOL04}). These new observations
  provide additional spectral information and allow us to investigate weaker
  spectral features, such as the two metallicity indices `1370' and `1425',
  defined by Leitherer et al. (\cite{LEI01}) and recently revisited by Rix et
  al. (\cite{RIX04}; herafter R04). These two indices measure the strength of selected absorption blends originating in the photospheres of hot stars. These indices are known to vary strongly with the metallicity. But they depend only weakly on the starburst ages and (originating from excited atomic states) they are not affected by interstellar absorption components. However, since the 1370 and 1425 indices are based on weak absorption blends in the vicinity of interstellar and wind lines, their measurement require, in addition to an adequate S/N, an adequate spectral resolution. The 1425 index has already been applied successfully to the lensed galaxy MS\,1512-cB58 ($z = 2.73$) by Leitherer et al. (\cite{LEI01}) and R04 and to Q1307-BM1163 ($z = 1.41$) by Steidel et al. (\cite{STE04}) and R04. The resulting metallicities were, respectively, $\approx 0.25\,\mathrm{Z}_{\odot}$ and $\approx 0.8\,\mathrm{Z}_{\odot}$. Moreover, de Mello et al. (\cite{DeM04}) used the 1425 index to derive an average metallicity of $\geq 1.5\,\mathrm{Z}_{\odot}$ for a co-added spectrum 
of five NIR-luminous galaxies with $\langle z \rangle \approx 2$ from the K20 survey (Cimatti et al. \cite{CIM02}; Mignoli et al. \cite{MIG05}). 

In Sect.~\ref{sec:Obsanddata} we describe the sample and the observations. In Sect.~\ref{sec:Prop} of this paper we report some basic properties of the observed absorption spectra. In Sect.~\ref{sec:results_indices} we present our  results on the metallicities and the metallicity evolution. Sect.~\ref{sec:conclusion} lists our conclusions. Throughout the paper we assume for the synthetic spectra used for comparison purposes continuous star formation (1\,M$_{\odot}$\,yr$^{-1}$), a starburst age of 100\,Myr, and a Salpeter IMF ($\alpha_\mathrm{IMF} = 2.35$; Salpeter \cite{SAL55})\, with masses between 1\,M$_{\odot}$ and 100\,M$_{\odot}$. This is a reasonable choice, since models with continuous star formation usually reproduce the 
rest-frame UV spectra of LBGs quite well and become insensitive to the age for $> 100$\,Myr. Moreover, as shown by Rix et al. (\cite{RIX04}), models with significantly different IMF slope or mass cuts would produce spectral features, which are not observed in typical LBG spectra. 

\section{Sample selection and observations}\label{sec:Obsanddata}
\subsection{Target selection for medium resolution spectroscopy}
All targets for our medium resolution spectroscopic program were selected from
the FDF spectroscopic survey of Noll et al. (\cite{NOL04}).  In this
  survey Noll et al. (\cite{NOL04}) observed the (observer's frames)
  $I$-brightest FDF objects for which there were photometric redshifts
  (Gabasch et al. \cite{GAB04}) available. In detail we selected galaxies
  which from the low-resolution ($R \approx 200$) spectra of the survey were
  known to have either bright restframe UV continua (to allow absorption line
  studies, as discussed in this paper) or strong Ly$\alpha$ emission (allowing
  an analysis of their Ly$\alpha$ profiles; cf. Tapken et al. \cite{TAP04},
  Tapken \cite{TAP05}).
\subsection{Observations and data reduction} 
The observations were carried out using the FORS2 instrument at ESO's Very Large Telescope at Paranal, Chile, with the holographic grisms 1400V and/or 1200R. Both grisms resulted in a spectral resolution $R \approx 2100$. The observed wavelength ranges depended on the targets' positions in the focal plane and are listed in Table~\ref{tab:obs}. For galaxies with $z \approx 2.4$ and $z \approx 3.3$ the C\,IV resonance doublet appears near the center of the 1400V and 1200R grism spectra, respectively. The restframe resolution element of the spectra is about $\approx 0.75$\,\AA{}, which matches the resolution of the synthetic spectra of Leitherer et al. (1999, 2001) which we used for comparison. All data were collected in service mode using for each grism one single MXU mask. The integrated exposure times, the average seeing values, and the number of objects observed simultaneously for the 1400V and 1200R grisms were, respectively, $6.25$\,h and $9.45$\,h, $0.81''$ and $0.92''$, and $n = 22$ and $n = 27$. Six objects were included in both setups. Thus, in total we observed 43 individual galaxies. The raw data were reduced using the MIDAS-based FORS pipeline developed by Noll et al. (\cite{NOL04}) and Tapken (\cite{TAP05}). Since the efficiency of holographic grisms strongly varies with the angle of incidence (and thus with the object's position  in the telescopes focal plane) no direct flux calibration of the medium resolution spectra was attempted. Instead we converted the relative flux scale of the medium resolution spectra to absolute fluxes using the well calibrated low-resolution spectra of the same targets published by Noll et al. (\cite{NOL04}). 
\begin{figure*}[ht]
\includegraphics[width=18cm]{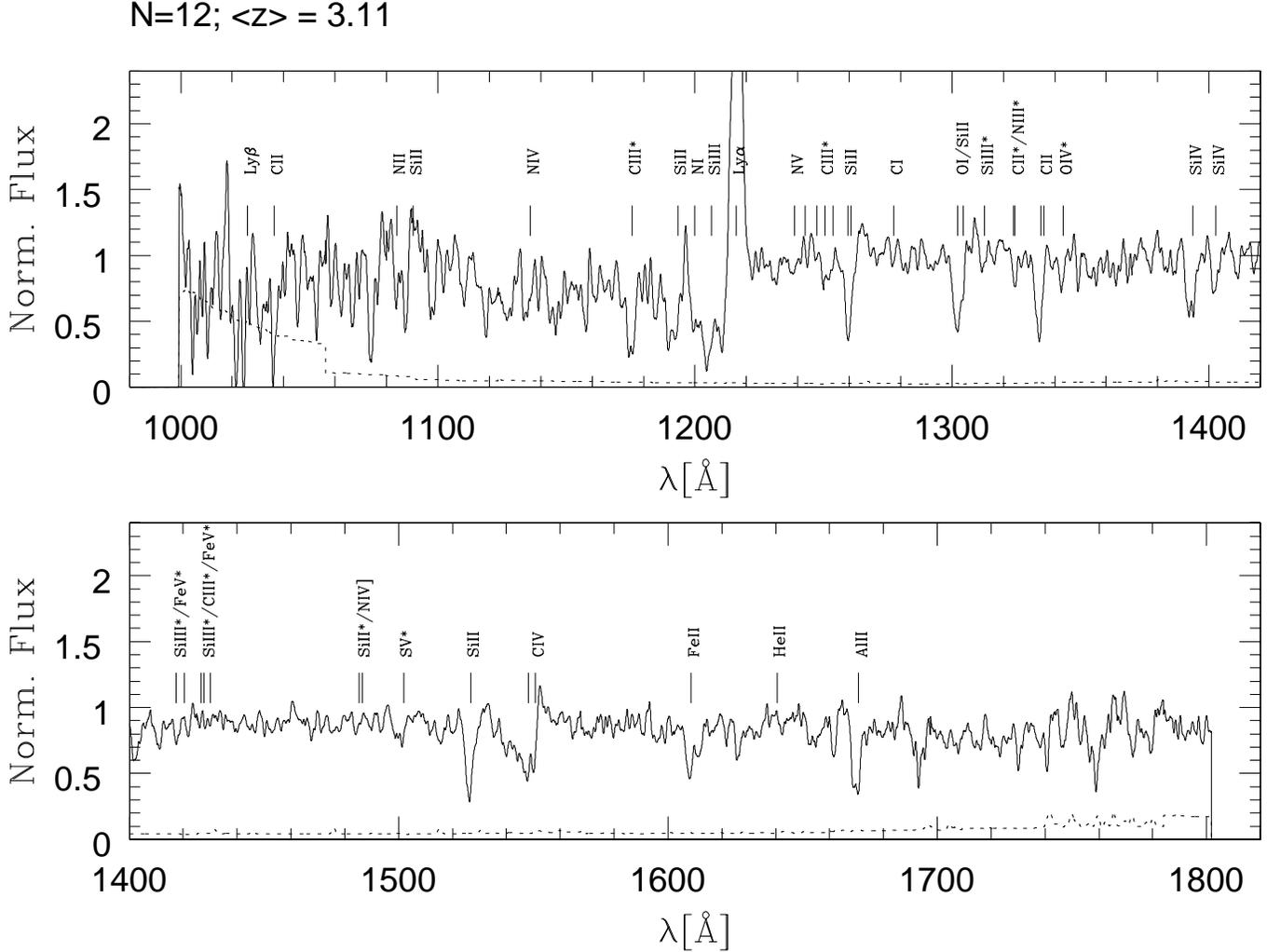}
\vskip.2in
\caption{Normalized F$_{\lambda}$ medium-resolution composite spectrum of 12 galaxies with $2.4 \leq z \leq 3.4$, $\langle z \rangle = 3.1$. The dotted line indicates the noise level. Some prominent spectral features are indicated by vertical lines. Purely photospheric lines are indicated by asterisks.}
\label{fig:mall_pap}
\end{figure*}
\begin{figure*}[ht]
\includegraphics[width=18cm]{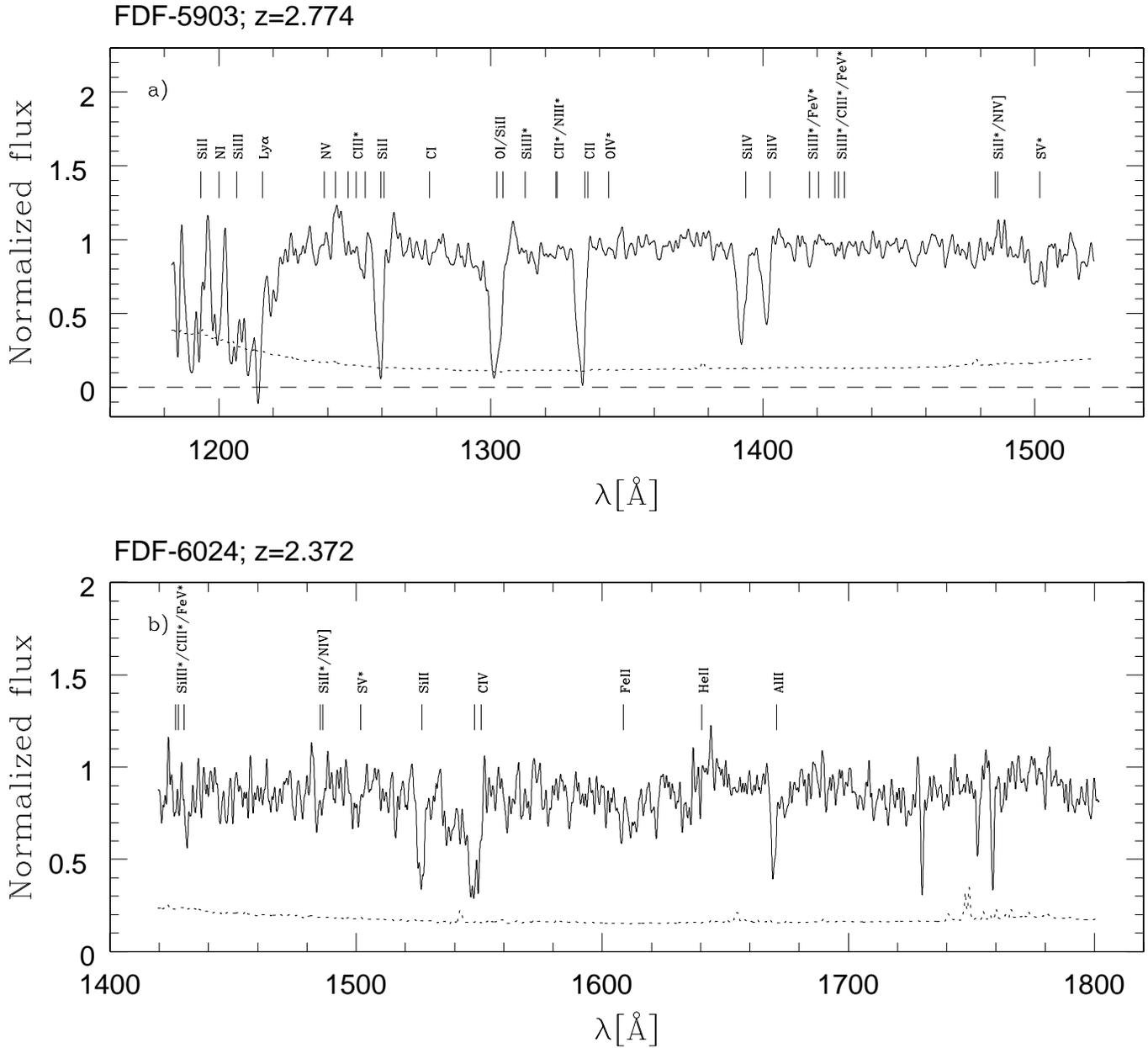}
\vskip.2in
\caption{Restframe spectra of FDF-5903 (a) and 6024 (b). The dotted lines indicate the noise level. A flux of unity corresponds to $10.83 \times 10^{-21}$\,W\,m$^{-2}$\,\AA$^{-1}$ and $11.60 \times 10^{-21}$\,W\,m$^{-2}$\,\AA$^{-1}$ for FDF-5903 and FDF-6024, respectively. Some prominent spectral features are indicated by vertical lines. Purely photospheric lines are indicated by asterisks. The strong absorption features between 1720 -- 1760\,\AA{} in the spectrum of FDF-6024 are due to a foreground metal absorption system at $z = 1.4885$.}
\label{fig:5903_6024}
\end{figure*}
\subsection{Sample selection}\label{sec:sample}
Many of the galaxies observed for studying the Ly$\alpha$ emission line profiles showed only weak continua and, therefore, were not suitable for investigating the absorption spectra. Hence, for the present paper we selected a subsample of 12 galaxies which had (in at least one of the two grisms) a continuum S/N~$\geq 3.5$. These galaxies are listed in Table~\ref{tab:obs}, which gives the following information: Designation (ID) from the FDF photometric catalog (Heidt et al. \cite{HEI03}), total Vega magnitudes for $R$ and $K_s$, redshift $z$ from Noll et al. (\cite{NOL04}), the grism(s) used, the observed wavelength range(s), the average continuum signal-to-noise per resolution element, the Ly$\alpha$ equivalent width, and the continuum slope parameter $\beta$ as defined in Noll et al. (\cite{NOL04}). In eight of the galaxy spectra of Table~\ref{tab:obs} we also measured individual 1370 and 1425 indices (see Sect.~\ref{sec:results_indices}). These individual values for these eight galaxies are listed in Table~\ref{tab:indices}.   The $z > 3$  galaxies of Table~\ref{tab:obs} have (within the statistical uncertainties) typical average restframe colours, Ly$\alpha$ emission, absorption line strength, and continuum slopes as observed in larger samples of high-redshift galaxies (e.g., Shapley et al. \cite{SHA03}; Noll et al. \cite{NOL04}, see also Table \ref{tab:sample_prop}). For $z < 3$ our medium resolution sample is too small for a statistically meaningful comparison, but the main properties (like Ly$\alpha$ emission absorption line strength and continuum slope) are again typical for this redshift.  The average $\beta$ for the eight $z > 3$ galaxies is $-1.98$, while the four z$<$ 3 galaxies are on average redder  with $\beta$ =  $-0.69$ in agreement with Noll et al. (\cite{NOL04}), who find that the galaxies with $2 < z < 3$ are indeed redder  than the galaxies with $3 < z < 4$. The eight galaxies used for determining the photospheric indices (see Table ~\ref{tab:indices}) also do not have unusual properties, except that the five $z > 3$ galaxies in Table ~\ref{tab:indices} have  weaker  Ly$\alpha$ emission as compared to the total FDF sample with $3 < z < 4$.
\begin{table}
\caption{Properties of the medium-resolution sample, which is subdivided into
  two redshift bins ($z < 3$ and $z > 3$). The table lists the number of
  objects in the redshift bin, the mean redshift, the Ly$\alpha$ equivalent
  width, the continuum slope $\beta$, the mean equivalent width $W_{\rm LIS}$
  of six low ionization interstellar absorption lines (LIS, see Noll et
  al. \cite{NOL04} for details) and the equivalent width of
  C\,IV\,\ll\,1548,\,1551. The redshift and continuum slope is taken from Noll
  et al. (\cite{NOL04}), the equivalent widths are measured in a corresponding
  composite spectrum of the low-resolution spectra. In addition the
  corresponding data for the total FDF sample is given in parentheses (see
  Table~3 in Noll et al. \cite{NOL04}).}
\label{tab:sample_prop}
\centering
\begin{tabular}{|c|c|c|}
\hline
$z$ range & $2 < z < 3$ & $3 < z < 4$ \\
\hline
$N$ & $4\, (42)$ & $8\, (22)$ \\
$z$ & $2.49\, (2.39)$  & $3.29\, (3.33)$ \\
$W_{\rm Ly\alpha}$ [\AA{}] & $7.12\, (6.90)$ & $-24.0\, (-10.2)$ \\
$\beta$ & $-0.69\, (-0.56)$ & $-1.98\, (-1.79)$ \\
$W_{\rm LIS}$ [\AA{}] & $ 1.92\, (1.98)$ & $1.36\, (1.55)$ \\
$W_{\rm C\,IV}$ [\AA{}] & $4.38\, (3.80)$ & $2.34\, (2.33)$ \\
\hline
\end{tabular}
\end{table}

 All the galaxies with $2.5 < z < 3.5$ in our medium resolution sample have colours (see Heidt et al. \cite{HEI03}) fulfilling the Lyman-Break criteria of Steidel et al. (\cite{STE95}, \cite{STE03}). The average $R$ magnitude of these galaxies is about $0.7$ brighter than for the $z \sim 3$ sample of Steidel et al. (\cite{STE03}). Hence, our $z > 2.5$ galaxies belong to the most luminous LBGs at $z \sim 3$. The colours of our galaxy subsample with $2 < z < 2.5$ match the selection criteria for LBG-like `BX' galaxies at $\langle z \rangle = 2.2$ (Steidel et al. \cite{STE04}). However, the average $R$ magnitude of our medium resolution galaxy subsample with $2 < z < 2.5$ is about $1.3$ brighter than for an average `BX' galaxy of Steidel et al. (\cite{STE04}). Two out of three galaxies  with $2 < z < 2.5$ show $K_s < 20$ (see Table~\ref{tab:obs}). Consequently,  the properties of these two galaxies overlap those of the Shapley et al. (\cite{SHA04}) sample. The seven $K$-bright galaxies of Shapley et al. (\cite{SHA04}) are characterized by high stellar masses in the order of $10^{11}\,\mathrm{M}_{\odot}$ and metallicities close the solar value. Our $z < 3$ galaxies show $J - K_s$ well below the criterion of  distant red galaxies defined by $J - K_s > 2.3$ (Franx et al. \cite{FRA03}; van Dokkum et al. \cite{DOK04}), implying that their restframe optical spectra do not have strong 4000\,\AA{} breaks.   
\subsection{Composite spectra}
Averaging all 12 spectra listed in Table~\ref{tab:obs} we produced and investigated a composite spectrum corresponding to a mean redshift of $\langle z \rangle = 3.1$. This composite spectrum is plotted in Fig.~\ref{fig:mall_pap}. Before co-adding the spectra all wavelengths were converted to the restframe scale using the individual redshifts derived by Noll et al. (\cite{NOL04}) and normalized to same flux F$_{\lambda}$ at 1280\,\AA{}. The noise level of the composite spectrum was estimated from the noise of the individual spectra as well as from the differences between the individual spectra. In addition, we calculated composite spectra of subsamples containing about half of the available data each. Only features which showed up in all the subsamples were regarded as general features of the galaxy spectra. Since the individual spectra cover different restframe wavelength ranges and since the spectra were added with equal weight, the S/N of the composite spectrum varies with wavelength with a mean value of $\approx 16$. Thus, the S/N of the composite spectrum is is not much higher than the S/N of 
the best individual spectra. However, in view of the differences between the individual spectra discussed below, the composite spectra should be more representative for the typical properties of high-redshift galaxies. 

Galaxy spectra of our redshift range are known to contain absorption features of intervening intergalactic clouds and galaxies. Such features are also present in our spectra. No attempt was made to remove these features before the co-addition. Thus, these features add somewhat to the noise of the composite spectrum. But, occurring at different restframe wavelengths, they are much diluted in the composite spectra. An exception is a strong metal absorption system at $z = 1.4885$ in the spectrum of FDF-6024. The strong Fe\,II (UV\,2 and 3) lines (EW$_{0} \leq 1$\,\AA{}) of this system cause spurious absorption features at the red end of the composite spectrum where only three of the 12 spectra contribute.       
\begin{figure}[ht]
\includegraphics[width=6.0cm,angle=-90]{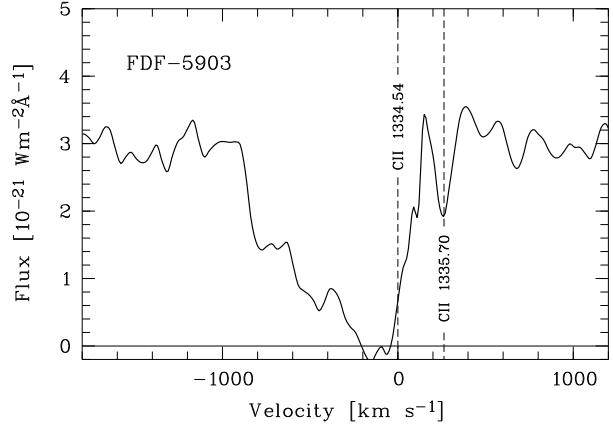}
\vskip.2in
\caption{Profile of the C\,II~1334.5\,\AA{} resonance line originating
in the interstellar medium of the galaxy FDF-5903. The vertical lines indicate 
the positions of the unshifted C\,II~1334.5\,\AA{} resonance line and of the excited 1335.7~\AA{} line of the same multiplet. The abscissa gives the velocity of the components of the 1334.5\,\AA{} line relative to the mean velocity of the photospheric absorption features. Note the complex structure and extended red wing of the resonance line.}\label{fig:fdf5903_c2}
\end{figure}

\section{Spectral properties of the observed high-redshift galaxies}
\label{sec:Prop}
\subsection{Overview}
All observed spectra show apart from strong differences in the Ly$\alpha$ line profile (ranging from pure absorption to strong emission) very similar starburst spectra. Most spectral properties are qualitatively in good agreement with those reported earlier from lower resolution composite spectra of this redshift range (see, e.g., Shapley et al. \cite{SHA03}; Noll et al. \cite{NOL04}). Since there exist only few spectra of high-redshift starburst galaxies with a spectral resolution comparable with the present data, we used our composite spectrum (Fig.~\ref{fig:mall_pap}) and the individual spectra of FDF-5903 at $z = 2.774$ and FDF-6024 at $z = 2.372$ (Fig.~\ref{fig:5903_6024}) to derive some characteristic properties of selected absorption lines for our sample of high-redshift galaxies. For the two galaxies FDF-5903 and FDF-6024 with a continuum S/N~$> 10$ we investigated the strength and profiles of selected stellar and interstellar absorption lines in the individual spectra. Results based on the composite spectrum for lines and blends originating from the stellar photospheres are listed in Table~\ref{tab:photosp}. The corresponding results pertaining to interstellar medium lines can be found in Table~\ref{tab:ism}. While the lines originating from excited energy levels listed in Table~\ref{tab:photosp} can be expected to be of purely photospheric origin, the resonance lines listed in Table~\ref{tab:ism} may contain in addition to the interstellar features photospheric and stellar wind contributions. 
\begin{table}
\caption{Photospheric stellar lines measured in the composite spectrum. 
Parentheses indicate lower accuracy measurements.}
\label{tab:photosp}\vspace{0.5cm} 
\centerline{\begin{tabular}{|l|c|r|r|l|}\hline
Ion & $\lambda _{\rm lab,vac}$ & EW$_{0}$ & $\Delta v$ \,\ \ \ \ & Remark \\
    & [\AA{}] & [\AA{}] \ & [km\,s$^{-1}$] & \\
\hline
C III & 1175.64 & 1.40 & ? & 6 lines \\
Si III & 1312.59 & .38 & -36 & \\
C II, NIII & 1323.93, 1324.32 & .33 & ? & blend \\
O IV & 1343.35 & .85 & -44 & blend \\
Si III & 1417.24 & .15 & (+254) & weak \\
Fe V & 1420.46 & .20 & ? & blend \\
S V & 1501.76 & .34 & +109 & \\
Fe III & 1531.60 & .10 & (-103) & weak \\
Fe III, Al III & 1611.74, 1611.87 & .29 & -27 & blend \\
\hline
\end{tabular}}                         
\end{table}              
\subsection{Equivalent widths of the interstellar lines}\label{sec:ism_ew}
Compared to MS\,1512-cB58 (Pettini et al. \cite{PET00}, \cite{PET02}) the interstellar lines are systematically weaker in our composite spectrum. The mean restframe equivalent width EW$_{0}$ of the strong low-ionization interstellar (LIS) lines Si\,II\,\lam\,1260, O\,I\,\lam\,1302 + Si\,II\,\lam\,1304, C\,II\,\lam\,1335, Si\,II\,\lam\,1527, Fe\,II\,\lam\,1608, and Al\,II\,\lam\,1671 of our composite spectrum is $1.72$\,\AA{}. This number is in good agreement with the value of $1.77$\,\AA{} and $1.54$\,\AA{} for redshifts of $\langle z \rangle = 3.1$ and $\langle z \rangle \approx 3$ derived by Noll et al. (\cite{NOL04}) and by Shapley et al. (\cite{SHA03}), respectively, from low-resolution composite spectra of large samples.
\footnote{In Noll et al. (\cite{NOL04}) it is shown that the existing redshift dependence is dominated by differences in the structure and kinematics of the interstellar medium at different epochs.} 
On the other hand, $1.72$\,\AA{} is only 60\% of the corresponding value ($2.90$\,\AA{}) for MS\,1512-cB58 (cf. Pettini et al. \cite{PET00}). Similar ratios between our sample and cB58 are also found for the weak LIS lines and for the higher ionization lines, such as Si\,III and Si\,IV. Although, as shown by Noll et al. (\cite{NOL04}), the LIS-EW$_{0}$ decreases with increasing redshift and increasing Ly$\alpha$ emission, the difference of the LIS strength between our spectra and MS\,1512-cB58 can only in part be explained by the higher mean redshift and stronger mean Ly$\alpha$ emission of our galaxies. Higher LIS equivalent widths are observed in the individual spectrum of FDF-5903. However even in this case the observed LIS lines reach only about 92\% of the cB58 
EW$_{0}$ values. Hence our comparison seems to indicate that the LIS lines are atypically strong in MS\,1512-cB58 (see also the discussion in Shapley et al. \cite{SHA03}). 

In our composite spectrum the observed equivalent width ratio of $\approx 1.4$ (cf. Table~\ref{tab:ism}) of the Si\,II lines at 1260 and 1527\,\AA{} (having very different $\lambda f$ values) indicates a high average degree of saturation of the LIS lines. Although the somewhat higher EW$_{0}$ of the 1260\,\AA{} line (having the higher $\lambda f$ value) may show that the saturation is not quite complete in some objects of the sample, the ratio of the equivalent widths is too close to unity to explain the large rest flux in both lines in Fig.~\ref{fig:mall_pap} by optical depth effects. Hence our medium resolution spectra support the suggestions by Steidel et al. 
(\cite{STE01}) and Shapley et al. (\cite{SHA03}) that differences in the H\,I gas covering factors rather than differences in the degree of saturation or abundance differences are the cause of the observed large variations of the LIS line strengths of high-redshift starburst galaxies.

 Since all LIS line profiles are well resolved at our spectral resolution, our spectra also provide new information on the optical depth and covering factors of the absorbing cool gas. A comparison of Figs.~\ref{fig:mall_pap} and \ref{fig:5903_6024} shows that in the spectrum of FDF-5903 the strongest LIS lines reach zero flux level in their cores, while this is not the case for the composite spectrum. From an inspection of the individual spectra we find `black' cores of the strong LIS lines for about half our sample, while the remaining objects (including FDF-6024) show significant rest flux even at the strongest observed LIS lines. Moreover, our new results also confirm that the covering factors for high-redshift galaxies are, on average, smaller than found by Heckman et al. (\cite{HEC01}) for local starburst galaxies. As pointed out by Steidel et al. (\cite{STE01}) this may have important consequences for the UV radiation field at high redshift.
\subsection{Kinematics of the interstellar lines}
The (instrumental-profile corrected) mean FWHM line width of the unblended strong LIS lines of our medium-resolution composite ($550 \pm 20$\,km\,s$^{-1}$) agrees well with the corresponding value ($560 \pm 150$\,km\,s$^{-1}$) found by Shapley et al. (\cite{SHA03}) from their low-resolution composite spectrum. All velocities listed in Tables~\ref{tab:photosp} and \ref{tab:ism} are given relative to the mean velocity of the four well measured photospheric lines listed in Table~\ref{tab:photosp}.  From the noise level of the medium-resolution 
spectra we estimate mean errors of the velocities and equivalent widths in Tables~\ref{tab:photosp} and \ref{tab:ism} to be about 50\,km\,s$^{-1}$ and $0.1$\,\AA{}, respectively. However, since the UV spectrum of hot stars is very rich, most features listed in Tables~\ref{tab:photosp} and \ref{tab:ism} 
are blends as shown, for example, in de Mello et al. (\cite{DeM00}). This may introduce additional errors which are difficult to estimate. We assumed for C\,IV that the sharp component at the red edge of the profile corresponds to the velocity of the interstellar contribution, but no attempt was made to derive the equivalent width of this component.

The interstellar lines of the two individual galaxies show broader, more complex, and (relative to the stellar photospheric lines) more blueshifted profiles. These profiles are also more complex than those observed in the spectrum of MS\,1512-cB58 by Pettini et al. (\cite{PET02}). This is illustrated by Fig.~\ref{fig:fdf5903_c2} where we plot the profile of the essentially unblended C\,II\,\lam\,1335 line of FDF-5903. As shown by this figure the interstellar lines of FDF-5903 consists of at least four components, including a narrow absorption component close to the systemic velocity, a strong component at $\approx -156$\,km\,s$^{-1}$, and additional components at $\approx -443$\,km\,s$^{-1}$ and $\approx -750$\,km\,s$^{-1}$. In all interstellar lines of this galaxy the profile extends bluewards to $\approx -900$\,km\,s$^{-1}$ and the total width of the absorption profile is $\approx 1100$\,km\,s$^{-1}$. The interstellar lines in the spectrum of FDF-6024 show narrower profiles (total width about 430\,km\,s$^{-1}$) with only two obvious components, one corresponding to approximately the systemic velocity and one at $\approx -190$\,km\,s$^{-1}$. The profiles are again asymmetric and extend blueward to $\approx -400$\,km\,s$^{-1}$. In contrast to FDF-5903 where the component at $\approx -156$\,km\,s$^{-1}$ always dominates, in the spectrum of FDF-6024 the relative strengths of two interstellar line components appear to depend on the total line strength. For weaker lines (Fe\,II\,\lam\,1608, Si\,II\,\lam\,1527) the low-velocity component is dominant, while for strong lines (C\,IV resonance doublet, Al\,II\,\lam\,1671) the blueshifted components appear stronger.

The fact that the FWHM of the interstellar lines in the composite spectrum are comparable to those of FDF-6024 alone and significantly smaller than in the spectrum of FDF-5903 shows that the other 10 galaxies included in the composite have on average much narrower interstellar lines. This is confirmed qualitatively by an inspection of the individual spectra. Fig.~\ref{fig:mall_pap} also shows that  the interstellar lines of the composite have on average less pronounced blue wings. Hence, although evidence for galactic wind absorption components (clearly observed in the individual spectra of FDF-5903 and FDF-6024) is also present in the composite spectrum, the outflow velocities and strength of the galactic wind absorption appear to be in the full sample on average smaller than in the case of the two individual galaxies investigated. This may be related to the fact that the two galaxies FDF-5903 and FDF-6024 are (in terms of observed absolute UV flux) among the most luminous galaxies in the sample\footnote{While FDF-5903 is definitely the most luminous object in the sample, the absolute luminosity of FDF-6024 is somewhat uncertain since this galaxy appears to be lensed ---and thus amplified--- by the massive foreground galaxy FDF-5908.}.

A comparison of Tables~\ref{tab:photosp} and \ref{tab:ism} shows that for our full sample we find only a small ($-56 \pm 36$\,km\,s$^{-1}$) mean blue shift of the interstellar lines relative to the photospheric features (where the error is almost entirely caused by the scatter of the photospheric line velocities). In order to check whether this shift resulted from the inclusion of the two individual galaxies discussed above or whether the shift is a general property of the sample, we repeated the measurement for a subsample of 10 galaxies with FDF-5903 and FDF-6024 excluded, but obtained practically the same value. Hence a small blueshift of the interstellar lines appears to be a general feature of high-redshift starburst galaxies. However, the shift observed in our sample is obviously smaller than the corresponding value 
($-210 \pm 80$\,km\,s$^{-1}$) found by Pettini et al. (\cite{PET00}, \cite{PET02}) for MS\,1512-cB58. It is also smaller than (but within $3\,\sigma$ error limits still consistent with) the value of $-150 \pm 60$\,km\,s$^{-1}$ derived by Adelberger et al. (\cite{ADE03}) and Shapley et al. (\cite{SHA03}) from their low-resolution composite spectrum.

In view of the significant differences in the behaviour of the interstellar lines in different galaxies described above the varying results appear not unexpected. On the other hand, our results clearly demonstrate that results based on individual galaxies or too small samples may not be representative for all high-redshift objects. Differences concerning the intrinsic interstellar lines are are not surprising since these features are expected to be influenced by the composition and kinematics of the interstellar medium (ISM), the presence and strength of galactic winds, and by the viewing angles of the individual galaxies.  

In good agreement with Pettini et al. (\cite{PET00}, \cite{PET02}), Shapley et al. (\cite{SHA03}), and Noll et al. (\cite{NOL04}) we find large differences ($\leq 760$\,km\,s$^{-1}$) between the velocities of the Ly$\alpha$ emission peaks and the interstellar lines. However, since only part of our targets show Ly$\alpha$ emission our data provide little statistical information on this topic. 

\begin{table}
\caption{Interstellar absorption lines measured in the composite spectrum}
\label{tab:ism}
\vspace{0.5cm} 
\centerline{\begin{tabular}{|l|c|r|r|l|}
\hline
Ion  &  $\lambda _{\rm lab,vac}$  &  EW$_{0}$ & $\Delta v$ \,\ \ \ \ & Remark \\
& [\AA{}] & [\AA{}] \ & [km\,s$^{-1}$] & \\
\hline
C II & 1036.34 & 1.63 & -46 & \\
N II & 1083.99 & .89 & +29 & \\
Si II & 1090.42 & 1.25 & -131 & \\
Si II & 1193.29 & 1.17 & -170 & ID?\\
N I & 1199.97 & ? & -77 & cont.? \\
Si III & 1206.51 & ? & +106 & blend \\
H I & 1215.67 & ? & -139 & blend \\
Si II & 1260.42 & 2.16 & ? & blend \\
C I & 1277.46 & .28 & +5 & \\
O I & 1302.17 & 2.57 & +39 & EW: OI+SiII \\
Si II & 1304.37 & 2.57 & -30 & EW: OI+SiII \\
C II & 1334.53 & 1.97 & -37 & \\
Si IV & 1393.76 & ? & -75 & bl. st.wind \\
Si IV & 1402.77 & ? & -111 & bl. st.wind\\
Si II & 1526.71 & 1.55 & -8 & \\
C IV & 1548.20 & ? & -53 & bl. st.wind \\
C IV & 1550.77 & ? & -130 & bl. st.wind \\
Fe II & 1608.45 & .64 & -18 & \\
Al II & 1670.79 & 1.45 & -142 & \\
Ni II & 1741.55 & .44 & -68 & \\
\hline
\end{tabular}
}                         
\end{table}              
\section{Metallicities and metallicity evolution}\label{sec:results_indices}
\subsection{Measurement of the photospheric indices}
\begin{figure*}[ht]\includegraphics{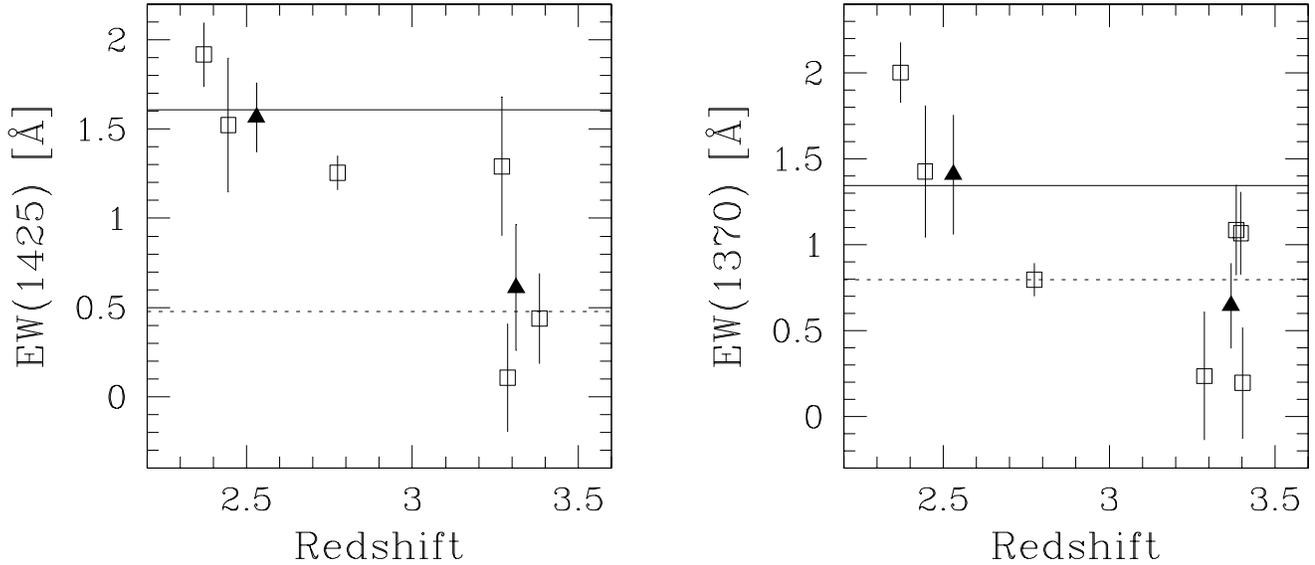}\vskip.2in
\caption{Measured metallicity indices 1370 and 1425 as a function of redshift. The open squares denote individual measurements. The filled triangles refer to mean values for the galaxies within the redshift bins $z \leq 3$ and $z > 3$. The horizontal lines correspond to the indices measured in the synthetic spectra of Leitherer et al. (\cite{LEI01}) for 1\,Z$_{\odot}$ (solid lines) and 0.25\,Z$_{\odot}$ (dotted lines).}
\label{fig:ews_z}
\end{figure*}

We measured the indices 1370 and 1425 in the spectra of eight of our target galaxies (see Table~\ref{tab:indices} and Fig.~\ref{fig:ews_z}). Since the two indices are defined as equivalent width values (in \AA{}) in the wavelength intervals 1415 -- 1435\,\AA{} and 1360 -- 1380\,\AA{}, respectively, their measurement depends critically on a correct ---or at least consistent--- derivation of the continuum level. Since there are only few ---if any--- regions in the UV spectrum of young stellar populations that are free of absorption features, the definition of the continuum is somewhat arbitrary. To be at least consistent with the data in the literature, we adopted for our EW measurements the 13 pseudo-continuum points defined by Rix et al. (\cite{RIX04}) in the wavelength interval between 1274\,\AA{} and 2113\,\AA{}. Using these pseudo-continuum points a model continuum was defined and the EW values were measured relative to this continuum by an automatic procedure. Normally the continuum level was determined from the low resolution (150I grism) spectra, which had been used for the flux calibration (see 
Sect.~\ref{sec:Obsanddata}) and which cover a wider wavelength range than the medium-resolution spectra. Only in cases where the low and medium resolution spectra showed differences, the continuum was estimated directly in the medium resolution spectrum. In these cases we either used pseudo-continuum points in the medium resolution spectra or we fixed the continuum interactively. For test purposes we applied these back-up procedure also to the objects where the continuum had been derived from the 150I spectra. A comparison of the results showed good agreement of the EW values derived with the different methods. The errors listed in Table~\ref{tab:indices} do not include possible systematic errors in the continuum determination (which are of the order of about 10 -- 15\%). These errors can produce an offset in Fig.~\ref{fig:ews_z}, but will not influence the slope of the relation plotted. In two cases the redshifted 1425 regions coincided with strong OH nightsky features. In these cases (indicated in Table~\ref{tab:indices}) a reliable measurement of the index 1425 was not possible.
\begin{table}
\caption{The observed indices 1370 and 1425 (in \AA{}).}\label{tab:indices}
\vspace{0.5cm} 
\centerline{\begin{tabular}{|l|c|c|c|r|}\hline
ID  &  $z$ & Index 1425 & Index 1370 & Remark \\\hline 
FDF-1337 & 3.403 & --              & $0.20 \pm 0.32$ & OH at 1425 \\
FDF-3173 & 3.270 & $1.29 \pm 0.39$ & --              & 1370 outside \\
FDF-3810 & 2.372 & $1.92 \pm 0.18$ & $2.00 \pm 0.18$ & \\
FDF-5550 & 3.383 & $0.44 \pm 0.25$ & $1.08 \pm 0.26$ & \\ 
FDF-5903 & 2.774 & $1.26 \pm 0.09$ & $0.80 \pm 0.09$ & \\
FDF-6063 & 3.397 & --              & $1.07 \pm 0.24$ & OH at 1425 \\
FDF-6934 & 2.445 & $1.52 \pm 0.37$ & $1.43 \pm 0.38$ & \\
FDF-7539 & 3.287 & $0.11 \pm 0.30$ & $0.24 \pm 0.37$ & \\
\hline\end{tabular}}                         
\end{table}              
In Fig.~\ref{fig:ews_z} we plotted the indices as a function of redshift of the corresponding galaxies. From the figure it is evident that both indices increase with decreasing redshift. A linear regression analysis resulted in a regression slope of dI/d$z$ = $-1.25$\,\AA{} for the index 1425 and 
$-1.00$\,\AA{} for the index 1370 with correlation coefficients of $0.828$ 
and $0.729$, respectively. 
\subsection{Comparison of the photospheric indices with theoretical models}In order to derive the metallicities corresponding to the indices 1425 and 1370 and to express the result of Fig.~\ref{fig:ews_z} in terms of chemical abundances, the two indices have to be calibrated. This can be done by measuring the indices in synthetic spectra ---like those of Leitherer et al. (\cite{LEI99}, \cite{LEI01}; SB99) or Rix et al. (\cite{RIX04}; R04)---  with known metallicity. These two sets of synthetic spectra SB99 and R04 differ in the use of different spectral libraries for constructing the model spectra. SB99 uses an {\it empirical} spectral library based on spectra of Galactic stars with solar metallicity observed with the International Ultraviolet Explorer (IUE) and Hubble Space Telescope spectra of $\approx 0.2\,\mathrm{Z}_{\odot}$ stars in the Large and Small Magellanic Clouds. To extend the libraries to higher and lower metallicities R04 included {\it computed} spectra based on the non-LTE line-blanketed model atmosphere code {\it WM-basic} of Pauldrach et al. (\cite{PAU01}). These theoretical libraries permit a free choice of metallicity. But in practice the evolutionary tracks and model atmospheres used for R04 were restricted to 2, 1, 0.4, 0.2, and $0.05\,\mathrm{Z}_{\odot}$. In order to calibrate our observed indices we used both sets of synthetic spectra. To be consistent we applied to the synthetic spectra exactly the same procedures as used for the observed spectra. Model spectra of higher resolution were smoothed to the resolution of the observed spectra before the measurements were made. Since both sets provide synthetic spectra for discrete metallicity values, we interpolated (or extrapolated) to the metallicities providing the optimal fits.  

As expected the different calibrations provide qualitatively consistent results. All metal indicators and both sets of models indicate a significant metallicity increase from subsolar mean metallicity at $z \approx 3.5$ to approximate solar metallicity at $z \approx 2.5$. In terms of cosmic time scales (for a universe with $\Omega _{\Lambda} = 0.7$, $\Omega _{M} =0.3$, $H_{0} = 71\,{\rm km\,s}^{-1}\,{\rm Mpc}^{-1}$) this metallicity increase detected in starburst galaxies would happen within $\approx 1$\,Gyr between cosmic ages of about $\approx 2$\,Gyr and $\approx 3$\,Gyr. However, there are quantitative differences between the results based on the two sets of synthetic spectra and there are inconsistencies between the metallicities derived from the different indices measured in the same set of synthetic spectra. For metallicities between the solar and the LMC/SMC value the index 1425 results in similar abundance values, but the two sets of model spectra result in different slopes of the index-metallicity relation. Thus, if the index 1425 is used the two model sets give different results for higher than solar as well as for lower than LMC metallicities. For the index 1370 the two model sets give a similar slope of the index-metallicity relation, but the two relations show a significant offset, SB99 giving metallicities which are about two times higher than those derived from the R04 models. It is outside the scope of the present paper to investigate the cause of these discrepancies, but it seems likely that subtle differences in the different model spectra are responsible. Future improvements in the model spectra may therefore well remove these differences.

In principle the measured indices and hence the metallicities derived also depend on the IMF of the starburst model. But as pointed out by Rix et al. (\cite{RIX04}), strong deviations from the Salpeter IMF would be accompanied by changes in other features like the C\,IV stellar wind line. Since these changes are not evident in our spectra, a significant departure from a Salpeter IMF appears unlikely. 

To derive metallicities from our spectra we applied the calibrations discussed above to the mean values of the photospheric indices 1370 and 1425 for $z \approx 2.5$  and $z \approx 3.3/3.4 $ (filled triangles in Fig.~\ref{fig:ews_z}). According to the discussion above the metallicities derived, which are listed in Table~\ref{tab:meanmet}, depend on the metallicity index and on the set of synthetic spectra used. E.g., for the epoch corresponding to $z \approx 3.3$ we find from index 1425 a metallicity of $Z/Z_{\odot} \approx 0.3$  and $\approx 0.2$ for the SB99 and R04 models, respectively, corresponding to an uncertainty of about 30\%. Using the index 1370 results in lower metallicities for both model spectra again with the higher value of $Z/Z_{\odot} \approx 0.2$ for SB99 and the lower one of $Z/Z_{\odot} \approx 0.1$ for R04. Hence, for this index the disagreement between the models is about a factor of 2. At the epoch corresponding to $z \approx 2.5$ the situation is more complicated: For the index 1370 the metallicity derived from the SB99 models ($Z/Z_{\odot} \approx 1.4$) is again higher than the one derived from the R04 models ($Z/Z_{\odot} \approx 0.6$). On the other hand the 1425 index gives a lower metallicity of $Z/Z_{\odot} \approx 1.1$ for SB99 and a higher one of $Z/Z_{\odot} \approx 1.7$ for R04. 
 \begin{table*}
\caption{Mean metallicities derived from the mean values of the photospheric 
indices 1425 and 1370 for $z \approx 2.5$  and $z \approx 3.3/3.4 $ (see
Fig.~\ref{fig:ews_z}) using the two sets of synthetic model spectra from SB99
and R04. For $z > 3$ the mean redshift for galaxies with measured 1425 index is $\langle z \rangle = 3.3$, for those with measured 1325 index $\langle z \rangle = 3.4$.}\label{tab:meanmet}
\vspace{0.5cm} 
\centerline{\begin{tabular}{|c||c|c||c|c|}
\hline
$\langle z \rangle$  & Z$(\langle 1425 \rangle)_{\rm SB99}$  & Z$(\langle 1425 \rangle)_{\rm R04}$ & Z$(\langle 1370 \rangle)_{\rm SB99}$ & Z$(\langle 1370 \rangle)_{\rm R04}$ \\
 & $[\mathrm{Z}_{\odot}]$ &$[\mathrm{Z}_{\odot}]$ &$[\mathrm{Z}_{\odot}]$ &$[\mathrm{Z}_{\odot}]$ \\
\hline
2.5     & 1.09 & 1.69 & 1.35 & 0.55 \\
3.3/3.4 & 0.33 & 0.19 & 0.16 & 0.09 \\
\hline
\end{tabular}}                         \end{table*}              
\subsection{Comparison with literature}
Comparing our new data with the two measurements of the 1425 index of high-redshift galaxies in the literature we note that for Q1307-BM1163 ($z = 1.4$), observed by Steidel et al. (\cite{STE04}), a metallicity of $0.7$\,Z$_{\odot}$ is obtained with both sets of model spectra, since this value falls into the metallicity range where (for this index) the index-metallicity relations of SB99 and R04 intersect. For the K20 sample (corresponding to about $z = 1.9$), investigated by de Mello et al. (\cite{DeM04}), the strength of the index 1425 ($2.3$\,\AA{}) would require a large extrapolation of the SB99 models (leading to an unrealistic metallicity of 12\,Z$_{\odot}$). The R04 models, which require a modest extrapolation only, result in $2.7$\,Z$_{\odot}$. As pointed out by de Mello et al. (\cite{DeM04}) the galaxies in the K20 sample are most likely the progenitors of local massive ellipticals. This may explain their high, super-solar metallicity. While this explanation is certainly plausible, we note that within the error limits the K20 data point would also be consistent with the metallicity evolution found for starburst galaxies. 
\section{Conclusion}\label{sec:conclusion}
We obtained medium resolution spectra of the restframe UV spectrum of 12
 distant ($2.3 < z < 3.5$) starburst galaxies. Our new data show that even at
 medium spectral resolution the high-redshift galaxies in the FORS Deep Field
 show relatively normal starburst spectra, resembling those observed in the
 local universe. The spectra differ by their strength and profiles of the
 intrinsic interstellar lines. The most complex profile with pronounced
 blue-shifted galactic wind components is observed in the spectrum of the most
 luminous galaxy of the sample. The differences in the behaviour of the ISM
 absorption lines may be due to differences in the ISM properties, differences
 in the strength of the galactic winds, and different viewing angles.  

 An evaluation of the metallicity indices 1370 and 1425 indicates that the
  abundance of heavy elements is increasing between $z \approx 3.5$ and
  $z \approx 2.5$. However, this temporal metallicity evolution should be taken with
  caution since our selection technique does not select galaxies by their masses. Therefore, if galaxies masses increase
  with time through hierarchical merging (e.g. Drory et al. \cite{DRO05}), it
  is possible that the temporal metallicity evolution that we are finding
  would not be as significant if evaluated in a mass-limited
  sample. Therefore, studies of large mass-limited samples at different
  redshifts (e.g. Savaglio et al. \cite{SAV05}) are needed in order to
  properly address the metallicity evolution.

 Our sample of 12 galaxies is too small to derive firm general conclusions
  concerning the cosmic chemical evolution. But our data show that the
  metallicity indices 1370 and 1425 are well suited to investigate relative
  abundances in high-redshift starburst galaxies and that it would be
  worthwhile to extend the small number of medium-resolution data. On the other hand since indices calculated from different sets of model spectra of the same chemical abundances are found to depend on the models used, and since the two indices give discrepant results, their calibration in terms of an absolute metallicity remains uncertain.  Future improvements of the model spectra will hopefully result in a more accurate calibration of the 1370 and 1425 indices. 

\begin{acknowledgements}It is a pleasure to thank the ESO Paranal Observatory staff for carrying out for us the service mode observations that contributed to this paper, and Dr. Jochen Heidt for valuable comments and discussions.  Moreover we thank the anonymous referee for valuable comments, that improved the paper. This research was supported by the German Science Foundation (DFG) (Sonderforschungsbereich 439).
\end{acknowledgements}


\begin{thebibliography}{}
\bibitem[2003]{ADE03} 
Adelberger, K.L., Steidel, C.C., Shapley, A.E., et al. 2003, ApJ, 584, 45
\bibitem[2002]{CIM02}
Cimatti, A., Mignoli, M., Daddi, E., et al. 2002, A\&A, 392, 395
\bibitem[2004]{DeM04}
de Mello, D.F., Daddi, E., Renzini, A., et al. 2004, ApJ, 608, L29
\bibitem[2000]{DeM00} 
de Mello, D.F., Leitherer, C., Heckman, T.M. 2000, ApJ, 530, 251
\bibitem[2005]{DRO05} 
Drory, N., Salvato, M., Gabasch, A., et al. 2005, ApJ, 619, L131
\bibitem[2003]{FRA03}
Franx, M., Labb\'e, I., Rudnick, G., et al. 2003, ApJ, 587, L79
\bibitem[2002]{FRY02} 
Frye, B., Broadhurst, T., Ben\'itez, T., et al. 2002, ApJ, 568, 558
\bibitem[2004]{GAB04}
Gabasch, A., Bender, R., Seitz, S., et al. 2004, A\&A, 421, 41
\bibitem[2001]{HEC01} 
Heckman, T.M., Sembach, K.R., Meurer, G.R., et al. 2001, ApJ, 558, 56
\bibitem[1998]{HEC98} 
Heckman, T.M., Robert, C., Leitherer, C., et al. 1998, ApJ, 503, 646 
\bibitem[2003]{HEI03}
Heidt, J., Appenzeller, I. Gabasch, A., et al. 2003, A\&A, 398, 49
\bibitem[2000]{KOB00} 
Kobulnicky, H.A., Koo, D.C. 2000, ApJ, 545, 712 
\bibitem[2001]{LEI01} 
Leitherer, C., Leao, J.R.S., Heckman, T.M., et al. 2001, ApJ, 550, 724 (SB99)
\bibitem[1999]{LEI99}
Leitherer, C., Robert, C., Heckman, T.M., et al. 1999, ApJS, 99, 173 (SB99)
\bibitem[1997]{LOW97} 
Lowenthal, J.D., Koo, D.C., Guzman, R., et al. 1997, ApJ, 481, 673 (L97) 
\bibitem[2001]{MEH01} 
Mehlert, D., Seitz, S., Saglia, R.P., et al. 2001, A\&A, 379, 96 
\bibitem[2002]{MEH02}
Mehlert, D., Noll, S., Appenzeller, I., et al. 2002, A\&A, 393, 809
\bibitem[2005]{MIG05} 
Mignoli, M., Cimatti, A., Zamorani, G., et al. 2005, A\&A, 437, 883
\bibitem[2004]{NOL04} 
Noll, S., Mehlert, D., Appenzeller, I., et al. 2004, A\&A, 418, 885
\bibitem[1979]{PAG79} 
Pagel, B.E.J., Edmunds, M.G., Blackwell, D.E., et al. 1979, MNRAS, 189, 95
\bibitem[2001]{PAU01} 
Pauldrach, A.W.A., Hoffmann, T.L., Lennon, M., 2001, A\&A, 375, 161 
\bibitem[2002]{PEN02}
Pentericci, L., Fan, X., Rix, H.W., et al. 2002, AJ, 123, 2151
\bibitem[2005]{SAV05} 
Savaglio, S., Glazebrook, K., Le Borgne, D., et al. 2005, ApJ, 635, 260
\bibitem[2000]{PET00}
Pettini, M., Steidel, C.C., Adelberger, K.L., et al. 2000, ApJ, 528, 96 
\bibitem[2001]{PET01} 
Pettini, M., Shapley, A.E., Steidel, C.C., et al. 2001, ApJ, 554, 981
\bibitem[2002]{PET02} 
Pettini, M., Rix, S.A., Steidel, C.C., et al. 2002, ApJ, 569, 742 
\bibitem[2004]{RIX04}
Rix, S.A., Pettini, M., Leitherer, C., et al. 2004, ApJ, 615, 98 (R04)
\bibitem[2005]{RIX05} 
Rix, S.A., Pettini, M., Leitherer, C., et al. 2005, ApJ, 615, 98 (R04) in ``Starbursts: From 30 Doradus to Lyman Break Galaxies'', eds.: 
R. de Grijs and R.M. Gonz\'alez Delgado, Astrophysics \& Space Science 
Library, Vol. 329. Dordrecht: Springer, 2005, p.311
\bibitem[1955]{SAL55}
Salpeter, E.E. 1955, ApJ, 121, 161
\bibitem[2003]{SHA03} 
Shapley, A.E., Steidel, C.C., Pettini, M., et al. 2003, ApJ, 588, 65  
\bibitem[2004]{SHA04} 
Shapley, A.E., Erb, D.K., Pettini, M., et al. 2004, ApJ, 612, 108
\bibitem[1995]{STE95}
Steidel, C.C., Pettini, M.,  Hamilton, D. 1995, AJ, 110, 2519   
\bibitem[1996a]{STE96a} 
Steidel, C.C., Giavalisco, M., Pettini, M., et al. 1996a, ApJ, 462, L17 
\bibitem[1996b]{STE96b} 
Steidel, C.C., Giavalisco, M., Dickinson, M., et al. 1996b, AJ, 112, 352
\bibitem[2001]{STE01}
Steidel, C.C., Pettini, M., Adelberger, K.L., et al. 2001, ApJ, 546, 665
\bibitem[2003]{STE03}
Steidel, C.C., Adelberger, K.L., Shapley, A.E., et al. 2003, ApJ, 592, 728
\bibitem[2004]{STE04}
Steidel, C.C., Shapley, A.E., Pettini, M., et al. 2004, ApJ, 604, 534
\bibitem[1994]{STO94}
Storchi-Bergmann, T., Calzetti, D., Kinney, A.L., et al. 1994, ApJ, 429, 572
\bibitem[2004]{TAP04}
Tapken, C., Appenzeller, I., Mehlert., D., et al. 2004, A\&A, 416, L1
\bibitem[2005]{TAP05}
Tapken, C., ``Ly$\alpha$ emission galaxies in the FORS Deep Field'', Doctoral Thesis University of Heidelberg, 2005
\bibitem[2000]{TEP00} 
Teplitz, H.I., McLean, I.S., Becklin, E.E., et al. 2000, ApJ, 533, L65 
\bibitem[2004]{DOK04}
van Dokkum, P.G., Franx, M., F\"orster Schreiber, N.M., et al. 2004, ApJ, 611,703

\end{thebibliography}
\end{document}